\documentclass[manuscript]{acmart}

\usepackage{array}
\usepackage{wrapfig}  % Import the wrapfig package
\usepackage{graphicx} % For including images
\usepackage[english]{babel}
\usepackage{csquotes}
\usepackage{multirow}
\usepackage{ragged2e}
\usepackage{array}
\usepackage[normalem]{ulem}
\usepackage{longtable}
\usepackage{tablefootnote}
\usepackage{footnote}
\makesavenoteenv{tabular}
\makesavenoteenv{table}
\usepackage{blindtext} %This package generates automatic text
\usepackage{soul}
\usepackage{epigraph}
\usepackage{dirtytalk}
\usepackage{amsmath}
\usepackage{lscape}
\usepackage[false]{anonymous-acm}

\AtBeginDocument{%
  \providecommand\BibTeX{{%
    \normalfont B\kern-0.5em{\scshape i\kern-0.25em b}\kern-0.8em\TeX}}}

\begin{document}

%Regrettable: Building Tools to Detect Low-Quality Ads in An Online Labor Market for Freelance Work
\title{Data Enrichment Work and AI Labor in Latin America and the Caribbean}
%Data Enrichment Work and AI Labor in

\author{Gianna Williams\textsuperscript{*}}
\email{gianna.williams@northeastern.edu}
\affiliation{%
  \institution{Northeastern University}
  \country{USA}
}

\author{Maya De Los Santos\textsuperscript{*}}
\email{maya.delossantos@northeastern.edu}
\affiliation{%
  \institution{Northeastern University}
  \country{USA}
}

\author{Alexandra To}
\email{alexandra.to@northeastern.edu}
\affiliation{%
  \institution{Northeastern University}
  \country{USA}
}

\author{Saiph Savage}
\email{s.savage@northeastern.edu}
\affiliation{%
  \institution{Northeastern University \& Universidad Nacional Autónoma de México (UNAM)}
  \country{USA \& Mexico}
}

\authoranon{
\renewcommand{\shortauthors}{Williams, et al.}}

\begin{abstract} 
The global AI surge demands crowdworkers from diverse languages and cultures. They are pivotal in labeling data for enabling global AI systems. Despite global significance, research has primarily focused on understanding the perspectives and experiences of US and India crowdworkers, leaving a notable gap. To bridge this, we conducted a survey with 100 crowdworkers across 16 Latin American and Caribbean countries. We discovered that these workers exhibited pride and respect for their digital labor, with strong support and admiration from their families. Notably, crowd work was also seen as a stepping stone to financial and professional independence.  Surprisingly, despite wanting more connection, these workers also felt isolated from peers and doubtful of others' labor quality. They resisted collaboration and gender-based tools, valuing gender-neutrality. Our work advances HCI understanding of Latin American and Caribbean crowdwork, offering insights for digital resistance tools for the region.

\end{abstract}

\begin{CCSXML}
<ccs2012>
   <concept>
       <concept_id>10003120.10003121.10003129</concept_id>
       <concept_desc>Human-centered computing~Interactive systems and tools</concept_desc>
       <concept_significance>500</concept_significance>
       </concept>
   <concept>
       <concept_id>10003120.10003121.10003126</concept_id>
       <concept_desc>Human-centered computing~HCI theory, concepts and models</concept_desc>
       <concept_significance>300</concept_significance>
       </concept>
 </ccs2012>
\end{CCSXML}

\ccsdesc[500]{Human-centered computing~Interactive systems and tools}
\ccsdesc[300]{Human-centered computing~HCI theory, concepts and models}

\keywords{gig work, crowd work, job market, labor market, freelancers, Latin America, crowdworker}

\received{15 July 2023}

\maketitle

\section{Introduction}
Crowdsourcing refers to a practice where tasks, projects, or assignments are outsourced to a large and distributed group of people, often referred to as the ``crowd" or the ``crowdworkers" \cite{Day_2023}. The people posting the tasks are known as ``requesters''. Both crowdworkers and requesters can be located around the world \cite{Naderi2018}. Crowdworkers frequently perform small discrete tasks that collectively contribute to a larger project or goal \cite{Naderi2018, Day_2023}. For instance, they might annotate images or videos to enhance AI services', leading to improved experiences for end users \cite{Naderi2018,Day_2023}. Workers usually perform the tasks for monetary gain \cite{Naderi2018,Day_2023}. Some of these tasks can include annotating, labeling, or translating data \cite{Naderi2018}. The labor that crowdworkers conduct is known as ``crowd work''. Crowdsourcing platforms include sites like Toloka, Amazon Mechanical Turk, or Appen \cite{Naderi2018,Day_2023}

An interesting dynamic that has emerged in crowd work, is that most crowdworkers are in the Global South \cite{Posch_2022}, while most requesters are in the Global North \cite{Posch_2022}. This can lead to a significant cultural gap between the requester and the worker, influencing differing opinions on optimal work practices, the kind of technical support that crowdsourcing platforms should offer, and approaches to resolving issues. In addition to cultural divides, crowd work is known to be differently experienced across genders \cite{Posch_2022, 10.1145/2998181.2998327}. As an illustration, earlier research has demonstrated that female crowdworkers often receive task recommendations that diverge significantly from those given to their male counterparts \cite{Posch_2022, 10.1145/2998181.2998327, 10.1145/3491102.3501834}. Additionally, the nature of social relationships established by women differs from those formed by men \cite{doi:10.1177/1024529420905187}.  Likewise, recent research on crowd work has unveiled a concerning trend: the earnings generated by Indian women crowdworkers were not allocated to them directly but were channeled directly to their spouses \cite{10.1145/3491102.3501834}. This practice has hindered these workers from receiving appropriate compensation for their efforts. Recent research also reveals that crowd-workers' work practices can diverge from the platforms' interface expectations due to workers' cultural backgrounds, potentially impeding workers' productivity and leading to worse work experiences \cite{10.1145/2998181.2998327}. Part of the problem is that crowdsourcing platforms are often crafted with a ``Global North" perspective, which might not accommodate practices differing from the designers' known regional norms. Similarly, the designers of these platforms may have been unaware of the dynamics emerging in regions like India, where female crowdworkers were not duly compensated for their labor \cite{10.1145/3491102.3501834}. These blind spots could lead to the creation of technologies that fail to align with the diverse needs of workers in the Global South. Given these dynamics, comprehending the experiences of crowdworkers, especially in Global South regions, is crucial. However, existing research has mostly centered on crowdworkers in India \cite{10.1145/3491102.3501834}, leaving knowledge underrepresented about how crowdworkers from other Global South regions operate \cite{reynolds2023cima}. Understanding these populations is vital, considering the rapid growth of crowd work in these regions. Remarkably, a substantial part of Tesla's autonomous vehicle labeling, designed for the Global South, hails from Venezuela \cite{Jones_2021}. In light of this, it is key to spotlight the experiences of crowdworkers from across a broader spectrum of Global South regions beyond India.

To bridge this knowledge gap, this paper hones in on comprehending the challenges and opportunities of Latin American and Caribbean crowdworkers through a survey study. We specifically concentrate on crowdworkers operating within the Toloka platform, which facilitates recruiting and engaging workers from this particular region. Demographically in our quantitative data, we found that Latin American crowdworkers differed from other global counterparts in a few key ways. For example, compared to Indian crowdworkers: 1) their economic status was relatively higher, and 2) substantially more identified as single (as opposed to married). Thematically in our qualitative data, we found that participants experienced a sense of pride and respect in their crowd work, and they also garnered respect from their family members and friends. Again this differs from low-income Indian women crowdworkers who, while feeling more confidence and agency, also felt societal and familial embarrassment from performing crowd work \cite{10.1145/3491102.3501834}. Our study also underscored that crowd work served as an initial step toward financial independence for Latin American workers, a distinct dynamic from the situation of crowdworkers in the United States who often associated crowd work with precarious labor conditions \cite{gray2019ghost}. At the same time, we observed that Latin American crowdworkers desired connections yet concurrently felt the crowdsourcing platform inadvertently fostered a sense of alienation from fellow workers. The platform appeared to be optimized in a manner that discouraged collaborative efforts. We contend that these findings stress the necessity for tools and interfaces custom-tailored to support Latin American crowdworkers effectively. To finalize the paper, we delve into a discussion of the design implications stemming from our data, and we offer a glimpse into potential avenues for future research in this domain.

This paper offers distinct contributions to the HCI community by unraveling how individuals from Latin America and the Caribbean participate in crowd work, sharing their unique experiences and viewpoints. The HCI community stands to gain valuable insights from this research.

\section{Related Work}

This section delves into the unique struggles of crowdworkers in various regions of the Global South to emphasize the critical distinctions between their experiences. Next, we highlight the need for formal research regarding crowd work in Latin America and the Caribbean, as we know little about their personal experiences and challenges. To contextualize our focus on the Global South, we then discuss Western companies' exploitation of crowdworkers in the Global North and the impact of these same companies expanding their reach into the Global South, known as digital colonialism. Combining these topics motivates us by exposing knowledge gaps surrounding the challenges and opportunities of crowdworkers in Latin America and the Caribbean. 

Before continuing, we define our stances on gender, feminism, Latin America and the Caribbean, and the terms Global South/Global North:

\begin{itemize}
    \item \textit{Gender}: We, as researchers, understand and recognize that gender is a social construct, and there are many different gender expressions. We do not subscribe to a strict gender binary. However, in our results, there is a fifty-fifty split between men and women in data where participants self-identify. 
    \item \textit{Feminism}: We define feminism as equal opportunity for everyone regardless of gender. We subscribe to intersectional forms of feminism where women are liberated regardless of race, class, sexual orientation, or gender identity .   
    \item \textit{Latin America and the Caribbean}: We, as researchers, do not hold a monolithic view of Latin America and the Caribbean. We understand that all Latin American and Caribbean countries have their own cultures, practices, and values. 
     \item \textit  {Global South/Global North} The terms Global South and Global North began being used during the start of globalization, where scholaar  Carl Oglesby \cite{alma9942389160001401}, differentiated  upwardly mobile communities (Global North) and lower standing economies (Global South) to justify the exploitation of goods and services. However, we understand these global binaries reproduce harmful rhetoric of what is considered a "developing" , "upwardly mobile" or "under-developed" country. Through the use of a global binary, we see “the entangled histories of modernity, colonialism, and capitalism” \cite{Harding_2016} that is present within globalization. We use this terminology to acknowledge the disenfranchisement of crowdworkers based on their geographical location and the growing exploitation by the Global North when working with other countries. We do not support rhetoric that portrays countries of the Global South as inferior or underdeveloped.  \end{itemize}

\subsection{Problems Facing Crowdworkers Globally}
    Most crowd work research initially fixated on enhancing the speed and quality of crowdworkers' tasks \cite{little2010turkit,chilton2013cascade}. Regrettably, the individual experiences of these workers were largely disregarded \cite{gray2019ghost}. Only recently have researchers begun to conduct interviews with crowdworkers \cite{maryX}, offering insight into their identities and the challenges they endure. The delay is partly due to a more recent growing emphasis on studying crowdworkers' experiences within specific regions \cite{maryX} as they are not a uniform group \cite{gordon2022jury}. This crucial realization has made it a priority for researchers to better understand crowdworkers' identities and requirements. Awareness of workers' unique cultural differences is vital for designing improved tools and platforms tailored to their diverse needs.

Even with this forward step, studying crowdworkers and their experiences on a global scale still presents a challenge due to the multitude of terms used to describe these individuals and their roles. The terminology associated with crowdworkers has evolved, resulting in a myriad of names, including:
``Clickworkers,'' ``Turkers'' (referring to Amazon Mechanical Turk workers), ``Tolokers'' (referring to workers on Toloka), ``Digital workers,'' ``Crowdworkers,'' ``Annotators,'' ``Ghost workers''.

\subsubsection*{Global South Crowdworkers.} The majority of research delving into crowdworkers' experiences in the Global South has predominantly centered around India \cite{Wiggers_2021}. An ethnographic study conducted by Gray and Suri \cite{gray2019ghost} sheds light on how, in India, a subset of crowdworkers with disabilities chose this path because of the discrimination and inaccessible resources commonly present in conventional employment. Similarly, Gray and Suri's research also highlighted that Indian women have begun participating in these platforms to augment their household earnings through online work \cite{gray2019ghost}. These women frequently are forced to surrender their income to their husbands \cite{10.1145/3491102.3501834}, who also supervise their crowd work. This dynamic was not observed in the Global North (particularly the United States) \cite{martin2014being}. If we aspire to create tools that cater effectively to crowdworkers' needs, it becomes imperative to comprehend the intricate social dynamics that encompass their experiences. In the context of India, it may be crucial to consider how we can design crowdworker tools that function within the framework of a patriarchal society \cite{sultana2018design}. 

Only recently have we begun to hear the narratives of Global South workers beyond India. Shockingly, OpenAI, the entity behind ChatGPT, paid Kenyan workers less than \$2 per hour to enhance the safety of its chatbot \cite{Xiang_2023}. These workers were frequently exposed to distressing content, including explicit and graphic text about child sexual abuse, bestiality, murder, suicide, and torture.  Consequently, many Global South immigrants still find themselves drawn to crowd work as an alternative form of income \cite{mcdowell2015roepke}. Research has revealed this bias against international workers remains prevalent across a diverse range of economies and platforms \cite{lehdonvirta2014online}. For instance, Beerepoot and Lambregt's investigation \cite{beerepoot2015competition} exposed a striking pattern of discrimination against foreign workers in various online job postings. Numerous listings employed discriminatory language, such as: \emph{``This job is not for people from Bangladesh and Pakistan, and your bid would be rejected automatically if you are from any one of the mentioned countries}" Additionally, their research identified instances where certain nationalities were exclusively permitted to apply, as seen in the following: \emph{``Business to Business appointment setters needed: with previous calling experience Filipinos are preferred."} Lehdonvirta et al. \cite{lehdonvirta2014online} refer to this phenomenon of discriminating against certain nationalities as the ``liability of foreignness." In their examination of Upwork, they discovered foreign contractors not only attract fewer opportunities but, furthermore, receive lower compensation for the same type of work compared to their domestic counterparts. The low-wage challenge crowdworkers face in the Global South stems from their foreign status and insufficient training and inclusion within their online jobs. A three-year randomized trial undertaken by MIT and Innovations for Poverty Action sheds light on this issue \cite{Wiggers_2021}. The study focused on crowdworkers in Nairobi, Kenya, who were part of the crowdsourcing platform Sama. Those who received training and inclusion in the platform's hiring pool demonstrated lower unemployment rates and higher average monthly earnings than those who received training alone.

Recent studies on East Asian crowdworkers unveiled a striking ability among these workers to foster trust with their requesters, leading to platforms becoming almost obsolete. Research by Graham et al. \cite{graham2017risks,graham2017digital} spotlights how crowdworkers from Vietnam adeptly forged robust relationships with their clients. This rapport enabled them to transition towards a direct working arrangement, bypassing the platform as an intermediary and eliminating the service fees typically imposed. Consequently, the workers reported a remarkable increase in their annual earnings, soaring from \$20,000 to \$40,000. Through this shift, these crowdworkers perceived themselves as distancing from any inequitable practices imposed by foreign companies operating in Vietnam.

Nevertheless, not all work relationships within the context of crowd work in this region are positive. Recent research into Australian crowd work platforms operating in East Asia shed light on how these platforms foster a dynamic that makes workers feel akin to ``maids" – a term denoting lower-paid laborers perceived as less prestigious than their engineering counterparts who earn significantly higher wages \cite{Bogle_2022}. This disparity in recognition and compensation is a concerning issue built from various factors. One contributing aspect to this problem is the inherent opacity of crowd work systems \cite{gray2019ghost}. These platforms often operate behind the scenes, obscuring the workers' presence and role. This hidden nature of crowd work contributes to the sense of invisibility experienced by workers, which platforms exploit to rationalize lower pay and subpar treatment. This invisibility can serve as a shield for platforms, enabling them to justify unequal treatment and compensation based on an underlying assumption of workers being secondary contributors.

Studies considering crowd work within Latin America and the Caribbean are scarce. This disparity can be attributed, in part, to the fact that these workers have been less outspoken regarding their online employment, hindering their identification for research purposes. Researchers focused on Latin America and other Global South regions have often had to rely on direct platform engagement to connect with workers since very few platforms allow researchers to filter or locate workers based on nationality. Galperin and Greppi showed noteworthy insights in a study centered on Freelancer, the preeminent Spanish-speaking platform for accessing online freelance projects spanning disciplines such as web and graphic design, copywriting, and editing \cite{galperin2017geographical}. Their research uncovered workers from Latin America were 42 percent less likely to secure bids posted by employers in Spain, a nation boasting the highest wages in the sample. Moreover, they disclosed that crowdworkers from Spain experienced a wage increment of 16 percent compared to their Latin American counterparts. This discrepancy underscores the urgency of comprehending the experiences of Latin American and Caribbean crowdworkers to develop collective action tools to address potential discrimination on crowdsourcing platforms and pressure platforms to support curtailing wage disparities.

Recent news reports have also begun investigating the intricacies of work dynamics in Latin America. One noteworthy example is the emergence of studies revealing a remarkable source of crowdworkers in Latin America: Venezuela \cite{Medina_2023,Chen_2019}. Numerous Venezuelan freelancers have shared their experiences of joining crowdsourcing platforms, where they tirelessly engage around the clock to meet their financial needs \cite{Medina_2023}. Many of these diligent workers find themselves in a situation where they must constantly connect to the internet to secure tasks. This challenge arises from lucrative assignments becoming available at hours that do not align with their local time zones. Nevertheless, driven by the urgent necessity for income, individuals from Venezuela demonstrate an exceptional willingness to log in and contribute their efforts regardless of the hour.

Their determination is partly attributed to Venezuela's ongoing political challenges, prompting many individuals to emigrate to neighboring Latin American nations. However, upon their relocation, these individuals often encounter hurdles in securing traditional employment opportunities due to the absence of country-specific certifications or qualifications \cite{Medina_2023}. In this context, crowd work emerges as a primary livelihood for these individuals. As they navigate the limitations of their new environment, crowd work offers a flexible avenue to earn income without the barriers imposed by traditional employment prerequisites. This situation underscores the impact of socio-political factors on work dynamics in the region and illustrates crowd work as a vital economic lifeline for those facing such difficulties. This aspect has been further underscored via interviews with Venezuelan crowdworkers, in which they stress their servitude is not to a crowdsourcing platform or AI but to the broader Latin American system \cite{Medina_2023}. These crowdworkers from Latin America accentuate that their lack of job security and income stems from residing in a low-income region. Nevertheless, the inadequate labor safeguards within these Global South regions likely enable crowdsourcing platforms to operate within these areas and offer wages below the minimum standards. This phenomenon is exemplified by the situation in Kenya, where OpenAI engaged African workers in crowd labor for compensation of less than 2 dollars per hour \cite{Perrigo_2023}, illustrating the concerning trend.

The exploitation of workers in the Global South is a pressing issue that demands our attention. While journalists have begun to shed light on these situations, their lack of engagement with HCI literature has caused them to overlook crucial insights that could inform the development of effective tools for workers in these regions. Our research into the experiences of crowdworkers in Latin America and the Caribbean seeks to address this gap by providing much-needed data on the challenges faced by this vulnerable population. By taking a culturally aware and sensitive approach to developing public tools, we can alleviate the pressures placed on these workers and improve their quality of life. We believe that there is a pressing need to learn more about their experiences in order to create appropriate tools that address their unique needs.

\subsubsection{Colonialism and Crowd Work}

Digital colonialism is a form of neo-colonialism that refers to the exploitation of the Global South by the West through control of their digital ecosystems \cite{coleman2018digital}. It occurs when Big Tech companies profit by extracting data from users around the globe without their consent or fair compensation, all the while centralizing power by maintaining digital infrastructures in their respective countries, a prime example being the U.S. Due to these corporations' vast influence, they can expand their services to reach users worldwide, preventing local innovators and industries from flourishing and becoming competitors to those in the West. Big Tech companies perpetuate power imbalances and reinforce existing inequalities by creating and nurturing digital dependencies where they control where and how data is stored and used. Additionally, they make the software they develop proprietary, meaning users cannot view or modify programs for their benefit or protection. Each time these companies deploy a new product or service under the pretense of altruism, they reinvigorate the exploitative cycle and concentrate even more power for themselves \cite{kwet2019digital}. This process is prominent in crowd work, made evident through powerful Western companies and the workers they decide to target to provide them with data. In crowdsourcing, a substantial amount of work is conducted worldwide, with a large portion completed by crowdworkers from the Global South \cite{schmidt2019crowdsourced}. The tasks achieved by crowdworkers from developing countries are then utilized to benefit requesters and companies in developed economies \cite{kassi2018online}. For instance, Indian crowdworkers might be assigned to label images for AI services, which are then utilized by UK-based technological companies and startups \cite{gray2019ghost,lehdonvirta2022cloud}. This observation has prompted researchers to propose that crowd work perpetuates historical systems of production and exploitation from European colonialism \cite{posada2021coloniality,graham2017risks,soriano2019between}, as is described by the term digital colonialism. 

\section{Methods}
In this section, we present our survey design, recruitment procedures, and data analysis methods. 

\subsection{Survey Design}  To comprehend the experiences of individuals from Latin America and the Caribbean who work on the Toloka crowdsourcing platform, we developed two surveys (one in Spanish and one in English, the latter targeting regions primarily English-speaking, such as the Caribbean Islands and South America). Notably, one of the authors hails from Latin America, an additional author is proficient in Spanish, and the remaining are fluent in English. Each survey encompassed 73 questions, blending closed and open-ended formats. Utilizing a 5-point Likert scale (ranging from 1 - Completely Disagree to 5 - Completely Agree), participants rated their agreement with specific statements, some accompanied by open-response queries for elaboration. The survey was organized into different thematic sections. Some of the main themes included:\\
\textbf{Career.} This section focuses on how crowd work relates to participants' career and professional goals. For example, we ask how crowdworkers view their jobs, how their jobs relate to their long-term career goals, and about any barriers that might exist in fulfilling those
goals.\\
\textbf{Work History.} It captures participants' prior crowd work experience on Toloka or other crowdsourcing platforms and asks about any additional or current employment beyond crowd work. \\
 \textbf{Labor.} It explores the task types workers engage in and their preferred work locations (e.g., cyber cafes, libraries, home). It also inquires about the payment methods they receive for their work. \\
 \textbf{Demographics.} It  studies the more specific context of our participants' social, educational, and economic capital. It is important to note that in studying these aspects, we drew inspiration from the methodologies employed in the censuses of Latin American countries \cite{Censo}. This approach allowed us to craft questions that were more culturally and regionally relevant. For instance, mirroring Latin American countries census \cite{Censo}, we inquired about individuals' socio-economic status through housing-related questions. These questions encompassed whether respondents owned or rented their homes, the number of rooms, construction materials, access to basic utilities (such as water and electricity), and ownership of assets like cars or appliances. We also questioned them about the size and composition of their households, recognizing that larger households or those with numerous dependents might face distinct economic challenges compared to smaller ones. In essence, by aligning our metrics with those utilized in Latin American census methodologies, we gained valuable insights into the demographics of our study participants. To understand how this compares to the general population, we report income data alongside publicly available census data from Latin America and the Caribbean. Our appendix presents our survey.
 
\begin{wrapfigure}{r}{0.55\textwidth}
\centering
  \includegraphics[width=0.55\textwidth]{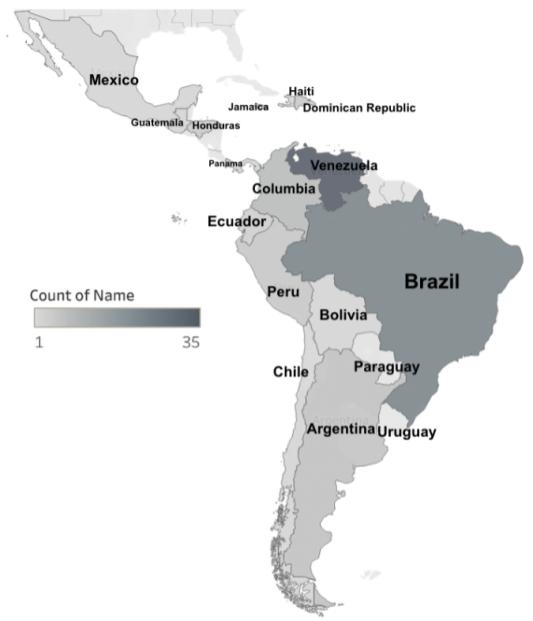}
  \caption{{A heatmap of participants' geographical location.}}
  \label{fig:mapFinal}
  \Description[A map of South America labeled by Country]{A map of  Latin American Countries with labels. Countries with high population count from our survey are in darker gray . Countries with low population is in lighter gray}
  \vspace{-0.5cm}
\end{wrapfigure}

\subsection{Recruitment}

To recruit participants for our study, we employed a stratified random sampling approach by country of residence (restricted to countries in Latin America and the Caribbean). After inviting crowdworkers from these countries to participate, we used Toloka's filters to recruit a proportionate stratified sample to ensure crowdworkers from any given country were not overrepresented in the data.

\subsection{Data Analysis} 
We used a mixed-method approach to analyze our survey data. First, we translated responses into English for better comprehension of the data, employing an Upwork college graduate person who focused on Latin American Spanish to English translation. All translations were cross-verified by the Spanish-speaking authors. This unified language aided our data analysis. 

For open-response data we used a grounded theory approach to qualitative data analysis. We began by open coding those survey responses along with study notes. The first author presented the open codes in a weekly research meeting for discussion and the research team iteratively distilled codes into a codebook. We then identified common experiences and connections across those open codes -- this yielded 10 axial codes which after further analysis resulted in 6 themes presented in the results. To complement our qualitative analysis, we identified the Likert scale questions linked to these main themes. Mean, median, and standard deviation was measured by the 5-point Likert scales (1=agree, 2=completely agree, 3=neutral, 4=disagree, 5=completely disagree). We report these results as relevant to the findings addressing our core research questions.

\section{Results}

In this section we will breakdown the results found in our study. We begin with the demographics of our particpants and then explore the themes found from our thematic analysis.  

\subsection{Participants}

100 crowdworkers from 16 countries across Latin America and the Caribbean participated in our survey study. The gender ratio was a 50:50 split between self-identified women and men crowdworkers. Notice that we do not endorse a rigid gender binary and allowed workers to open-response self-identify their gender.  Figure \ref{fig:mapFinal} displays a heat map showing the countries where our participants currently reside, all of whom are originally from those countries (i.e., country of residence and country of origin were the same for all participants). Most of our participants were from Venezuela (36\%) and Brazil (24\%). Our participants included a range of ages where 25 to 35-year-olds were the most represented (40\%), followed by the youngest group 18 to 25 (28\%,) and the oldest participant was 60 (1\%). The age demographics align with those found among crowdworkers in the United States and other Global South regions, where the average age typically ranges from 25 to 35 years \cite{Day_2023}. The crowdworkers in our study stated that (90\%) worked either twice a week or once a week on Toloka, dedicating most of their day to these tasks. Their work primarily involved activities such as translation (13\%), data labeling (35\%), Beta Testing (25\%), and completing surveys (25\%). In terms of education, participants held college degrees, with Bachelor's degrees at 31\% and Associate degrees at 13\%. Furthermore, 25\% of the participants had received technical vocational training.

\begin{figure}
    \centering
    \includegraphics [width=\textwidth]{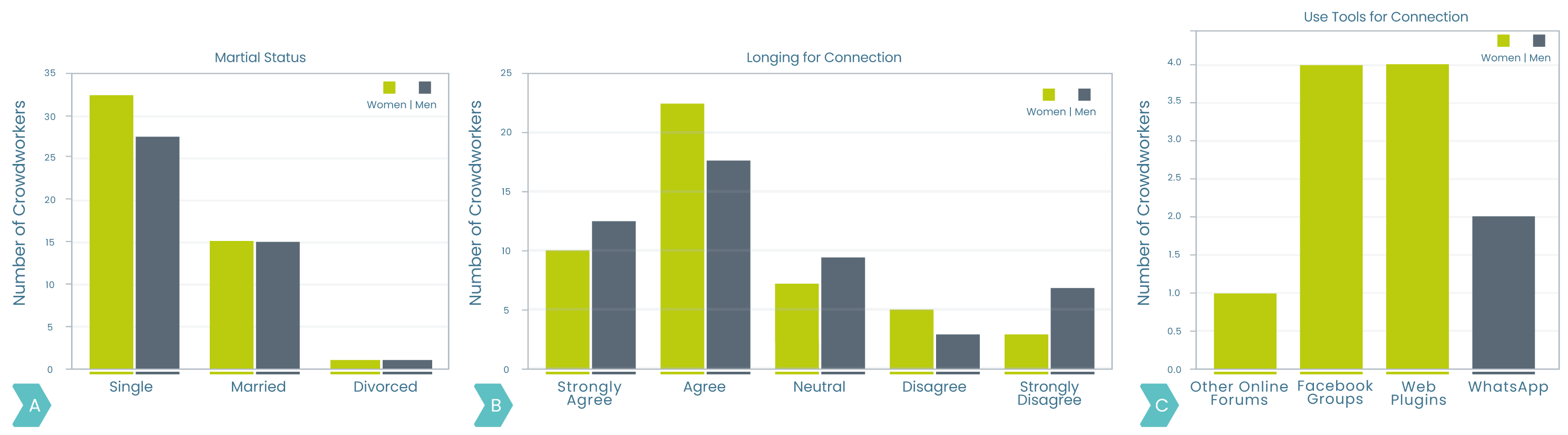}
    \caption{Overview of workers gender and their: A) marital status; B) attitudes towards seeking connections; C) usage of tools for connections.}
 \Description[Three colored bar graph]{This figure holds 3 separate graphs to represent gender dynamics within our study. Women are represented in light green and men are represented in gray.  Graph A with women represented in green and men in gray shows the martial status of participants. Graph B with participants split by gender (women in green , men in gray) to show how they feel about  wanting to connect with other crowdworkers. Graph C is split into gender (women in green , men in gray) where participants are asked if they use online tools to build better connection. 
    }
  
    \label{fig:genderstats}
    \centering
    
    \includegraphics [width=\textwidth]{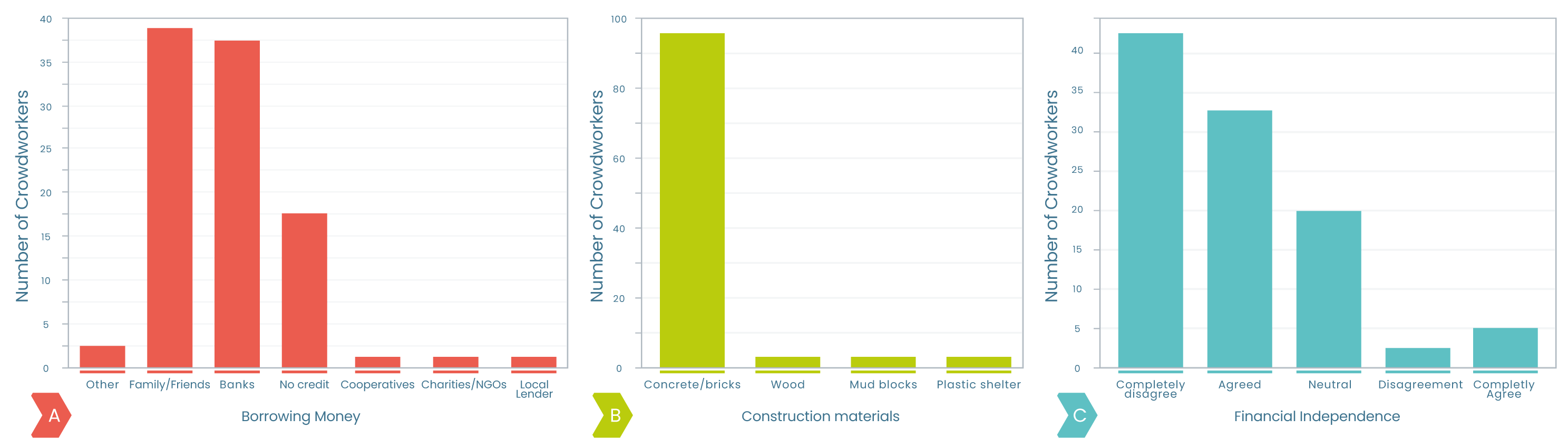}
    \caption{Overview of crowdworkers' financial practices: A) Illustrates the sources from which workers borrow money when in need; B) Depicts the predominant construction materials used in the homes of crowdworkers; C) Reflects workers' consensus on how crowd work contributes to their financial independence.}

        \Description[Three colored bar graph]{This figure holds 3 separate graphs to represent financial demographics of participants within our study. Graph A in red about how participants borrowed money , Graph B in light green represents construction materials used to build their homes. Graph C represents if they use crowd work for financial independence}
    \label{fig:finance}
\end{figure}

We also examined participants' marital status, their desire for connections with other workers, and the various tools they use for connecting (see Fig \ref{fig:genderstats}). Our interest in marital status stemmed from prior research indicating that a substantial number of women crowdworkers in the global South, like India, tend to be married, with their husbands often managing their platform earnings \cite{10.1145/3491102.3501834}. We studied participants' longing for connections and the tools they employ for this purpose by gender because earlier research has highlighted gender differences in this aspect \cite{eagly2007through,becker2002discrepancies}. We thus aimed to investigate if these differences were present in our study.

Fig \ref{fig:genderstats}.A presents the martial status of our participants. In difference to prior work \cite{10.1145/3491102.3501834}, our Latin American crowdworkers were single (59\%). Figure \ref{fig:genderstats}.B illustrates crowdworkers' desire for connections with their peers. Overall, both men and women expressed a longing to connect with other workers (64\% overall, 66\% women, 50\% men). Notably, despite the strong desire for connections, only 22\% of workers took steps to use technology for connecting with their peers (Figure \ref{fig:genderstats}.C). Furthermore, women tended to experiment with a wider range of technologies for this purpose, while men primarily relied on Whatsapp. In our survey, we also examined the socio-economic status of our participants, a topic of significance in prior research \cite{gray2019ghost,hara2019worker,hara2018data}. To understand more about the socio-economic backgrounds of our participants, we adopted a method similar to that used by Latin American countries in their census \cite{Censo,tahanyinfluencia}. Figure \ref{fig:finance} provides insights into the financial situations of our participants. Figure \ref{fig:finance}.A showcases the various sources from which participants feel comfortable borrowing money. Notably, banks were among the popular choices, alongside borrowing from friends and family. This finding aligns with the fact that Latin American countries often have lower trust in banks and rely on robust social networks for financial support \cite{evans2014empirical,donovan2012mobile,khan2016machine,malaquias2016empirical,suarez2016poor}. It is worth considering this result when contemplating the design of tools to assist crowdworkers in managing their finances. Perhaps incorporating a social component could be beneficial \cite{chiang2017understanding}.  

Another economic aspect we examined was the construction materials used in our participants' residences (Figure \ref{fig:finance}.B). Our analysis showed a predominant use of concrete bricks, indicating a possible middle-class socio-economic status \cite{padilla2017estudio,eduardomicrocreditos,tahanyinfluencia}. Responses to the other socio-economic questions hinted that the majority of our participants were middle-class. For example, all participants reported completing crowd work directly from their own computers, suggesting having funds for computer equipment, and they all had proficiency in English, another indicator of socio-economic status within Latin America, as not all public schools offer foreign language instruction \cite{madrid2010contexto,bori2019clase,meo2013habitus}.

We also studied the extent to which participants felt that crowd work contributed to their financial independence (Figure \ref{fig:finance}.C). This variable is significant because previous research has suggested that crowd work often leads to jobs with limited upward mobility \cite{10.1145/2441776.2441923,gray2019ghost}. Figure \ref{fig:finance}.C reveals that 73\%  participants either ``completely agreed'' or ``agreed'' that crowd work had indeed helped them achieve financial independence. It is worth noting that requesters, likely from the Global North, may pay higher wages to participants, contributing to this outcome \cite{kwet2019digital,coleman2018digital,posada2021coloniality}. This finding stands in contrast to prior research, which suggested that crowdworkers tended to originate from lower social classes. We saw the best way to define income within the context of Latin America is to ask participants about the status of their homes.  These results offer valuable insights into the demographic profile of crowdworkers in our specific context, emphasizing educational attainment and challenging previous assumptions about the socioeconomic backgrounds of crowd work participants and Latin America.  Before moving forward, however, we understand that 100 participants are not a representative sample of crowdworkers from Latin America. However, we believe this will create a discussion for new research to study Latin Americans within HCI. 

Next, we present the themes that emerged from participants' open-ended responses. Whenever possible, we establish connections with the Likert scale responses to provide more context on the experiences of crowdworkers.

\subsection{Worker Alienation and Lack of Worker Connections.}

Our survey revealed that  on a 5-point Likert scale, participants expressed ``Agreement''  (Mean= 3.58, SD=1.22, Median= 4)  with the need to establish connections with fellow crowdworkers. When asked to go into more detail, participants recognized the advantages of establishing connections with each other, especially to receive clarifications on task instructions:\emph{``It could help [connecting with other workers] to get clarifications on some of the task instructions'' (P21)}. They also considered it could help them to learn to navigate any new situation that emerges on crowdsourcing platforms: \emph{``Their expertise could greatly enhance my ability to effectively navigate any unfamiliar situations that I may encounter.'' (P55).} Some of the workers also felt these connections could provide them with access to alternative help that could surpass the capabilities of official platform support:\emph{``In the event of an issue arising, a fellow worker could offer an explanation, potentially leading to a quicker resolution compared to reaching out to technical support'' (P2)}.

Our participants thus demonstrated an awareness of the significance of connecting with each other. However, simultaneously, they exhibited feelings of ``alienation" from one another. When asking participants if they  ``{feel close to their coworkers on Toloka}'' the response on a 5-point Likert scale was ``Disagree'' (Mean = 2.71, SD = .96, Median=3), indicating that workers did not feel connected to one another. Workers perceived that the Toloka platform made it challenging to establish connections with others, and the platform did not seem conducive to building peer connections: \emph{``I can't find a clear way to do it [collaborate with others], I'm part of several Toloka crowdworker Facebook groups, but participants tend to only brag about the good tasks they found and their earnings, nobody foments collaborations, it is fictitious [that we collaborate]''} (P32). This result is consistent with previous research in crowd and gig work, which has shown that workers often experience feelings of isolation and lack of collaboration \cite{10.1145/2470654.2470742} brought on by the optimization of the platform. However, particularly interesting was that some of our participants expressed that they did  \emph{`` not interested in having contact with other workers'' (P52).} Some also held negative perceptions of their peers, which likely increased their alienation and deterred them from trying to connect with others. Specifically, there was a tendency to view other workers as ``foreigners" who were solely interested in ``easy money'' and were indifferent towards the quality of their work: ``\textit{They [other workers] are people from different countries, and most of them only care about trying to make a quick buck, regardless of the quality of their work} (P35).''
These feelings of alienation led some workers to push back against collaborations and argue that building connections was unnecessary and irrelevant, fostering competition. One participant believed: ``\textit{I do not feel that we should even worry about collaborations, because it has nothing to do with what we do at Toloka} (P72).'' The reluctance to engage with fellow workers could partly stem from past negative encounters while attempting to connect with the official support of the crowdsourcing platform and with requesters. Unfortunately, these attempts yielded no response, fostering an apprehension towards seeking connections with others: \emph{``I don't know about connecting with other workers. I know that there is a special panel where you can connect for help with Toloka and have communication with the requester, but they have never responded to my suggestions or observations'' (P65). } These findings hold significance because previous research has advocated for tools to facilitate worker connections \cite{10.1145/2702123.2702508}. However, it is key to consider the resistance and social dynamics that may exist when it comes to trying to initiate worker connections. It is key to take into account the multifaceted social dynamics that may exist.

\subsection{Respect for Crowd Work.}

From our survey, participants `agreed' (Mean = 3.95, SD = 1.23, Median = 4) that their family and friends held respect for the work they perform on crowdsourcing platforms:  this was further represented when participants said \emph{``My family likes seeing me work on crowdsourcing platforms'' (P26).} Recognizing the credibility of the work seemed to exert a positive attitude for the participants. By family members acknowledged the significance of crowd work family and friends allocating dedicated time and space for participants to engage in crowd work: \emph{``My family knows this job is real, and that I need my own time and space to complete it'' (P7).} P51 even mentioned how being a crowdworker provided her with a valid excuse to skip family events, granting her the dedicated time she needed to focus on her work: \emph{``They [my family] give me freedom to focus on my crowd work, and I can also be excused from family events'' (P51).}

The respect observed for crowd work seems to be closely linked to its tangible financial benefits: \emph{``They don't fully comprehend this new crowd work job, but they started to respect it more when I began earning similar to what I was making in my main job'' (P32)}. The substantial pay has played a pivotal role in elevating crowd work's status as a legitimate profession: \emph{``In the last months, my close family has seen that the crowdsourcing platforms really represent a strong source of income for me, and they have categorized it as a legitimate job. My friends also see it this way'' (P35).} Participants emphasized how their earned income not only contributed to their own prosperity but also positively impacted their social circle. This aspect appeared to play a role in garnering the respect others had for their work: \emph{``My family respects what I do. They know that the way I make money is through crowdsourcing platforms, and thanks to that I have the possibility of economically supporting our family expenses, and financially support my studies'' (P80)}. We also saw how participants used much of their earnings to support their families: \textit{"I spend the money I earn on the house. Whoever lives in the house has access to those benefits, such as the water and electricity bill, which is paid with my online earnings. So yes, the family does benefit from it."}(P4) This theme presents a notable contrast to prevailing research on crowd work in the United States \cite{strunk2023building}\cite{ASHFORD201823}. Unlike in the US, where gig work is often associated with feelings of shame or humiliation, in the context of Latin America, we observed a different dynamic. The family and friends of crowdworkers in this region showed respect for the nature of their labor.

\subsection{Crowd Work and Gender}

In the preliminary stages of our investigation, we posited that gender would exert a significant influence on crowdworkers' experiences. Our hypothesis was that women might encounter difficulties in utilizing the platform in comparison to men, owing to the historical backdrop of patriarchal norms \cite{HEILMAN2012113}, and to recent research that has uncovered how women gig workers tend to face greater challenges than their male counterparts \cite{kasliwal2020gender}. Hence, we were eager to grasp participants' viewpoints on how they perceived gender's impact on their experiences and outcomes.

However, we were surprised to find that  our participants  `disagreed' (Mean=2.10, SD=1.20, Median 2) when asked if their gender plays a role in the outcomes of crowd work. Expressing their views, one participant remarked, \textit{``These jobs don't need to consider gender. The same job can be performed by a man or a woman." (P40).} A prevailing perception that emerged among our participants was that everything was equal on crowdsourcing platforms, and gender did not show any influence on the outcomes. Any perception suggesting otherwise was perceived as lacking authenticity and regarded more as a fictional belief: \emph{``This type of work [crowd work] does not discriminate against genders, in that aspect, I personally consider that it [the role that gender plays in crowd work] is an ideological creation and it is not something that is real in this type of work'' (P35)}. Perceptions of crowd work as a neutral profession were shared by both men and women in our study. However we understand that growing research within crowd work has shown women in particular are more underpaid and under valued on crowd sourcing platforms based on how the algorithm sees gender. From this, we understand that participants in Latin America don't see 
a gender divide in their work. However, when asked question involving connection and financial independence more women agreed with these statement concepts (e.g., Mean=4.16, Median=4,, SD=.91). Henceforth  this showcases a gender gap around the need for connection and independence that should be explored.

\subsection{Crowd Work Powering Financial and Professional Independence.} 

Our participants `agreed' with the statement that they experience of financial independence seen in \ref{fig:finance}.C : \emph{``Thanks to crowd work I have an extra income and for the first time I can buy myself things without having to ask anyone [for money] (P33).''} ( Mean = 4.06, SD=1.06, Median=4)  Most of our participants used their earnings to contribute to family expenses and bills. However, crowd work was also seen as a source of income that bestowed upon them the ability to acquire things they personally desired, marking the first time they could do so for themselves. This newfound financial autonomy empowered them to fulfill personal aspirations and attain items of their choice, reflecting the transformative impact of crowd work on their ability to meet personal needs and desires: \emph{``I use the income I earn to buy food and also for pleasure outings (movies, restaurants)''(P56).} 

Some of the women in our study who identified as houseswives said crowd work granted them access to their own income and autonomy : \emph{``I'm a house wife […] at the moment Toloka is my only source of income, so it will surely help me to be able to become independent one day'' (P75).} Our research reveals a distinct trend in Latin America and the Caribbean, here we see  crowd work serving as a pathway to financial independence and the fulfillment of personal material aspirations. This pattern underscores the transformative role of crowd work in empowering individuals within the region to secure their financial well-being and achieve personal goals. Both men and women in our study also revealed that crowd work played a role in fostering their professional independence. A majority of the participants ( Mean = 1.6, SD=1.18, Median= 1) reported operating without any form of supervision during their crowd work engagements: 
\emph{``[In crowd work] I am the person who is directly responsible for my work, so I am the only one who is in charge and responsible for administrating and controlling the work I do'' (P78).} Notably,  crowdworkers reported "Completing Agreeing"  ( Mean = 4.52, SD=.86 ,Median=5)  with the notion of having time for self-care. As crowdworker P4 further shared: \emph{``Doing crowd work gives you a lot more flexibility. I don't have to take any type of public transportation to get to work, which consumes a lot of time, I could work 12 hours daily and still have time for sleep and time for myself'' (P4). } Some self-care activities workers enjoyed included sleep and following their favorite routines: \emph{``Crowd work gives me time to rest, to workout, and time for my beauty routine'' (P10)}, and most importantly, just having time for themselves: \emph{``Crowd work normally lets me distribute well my time, in the mornings I go out and do exercises, then I do home chores. Afterwards, I take some time to relax, and then I get to work on the crowdsourcing platforms'' (P34).} This aspect highlights the empowering nature of crowd work, particularly in the Latin America and Caribbean region, offering individuals the opportunity to exert control over their labor, their time, and advance their professional and personal goals. Remarkably, this result contrasts with findings in India \cite{10.1145/3491102.3501834}, where Indian women did not enjoy the same level of freedom in conducting crowd work. Instead, Indian women often worked under the supervision of their husbands, limiting their autonomy in this domain. This disparity underscores the significance of regional and cultural context in shaping the experiences of crowdworkers, highlighting the varying degrees of professional independence across different Global South regions.

\section{Discussion}

In this section, we present a discussion of the results and connect them to prior 
literature on the topic of crowd work. Our findings provide insight into the relationship between crowdworkers within Latin America and how Toloka treats these workers. Adding to the limited body of research on the experiences of  Latin Americans within crowd work we hope to bring more attention to the HCI community around the grievances facing Latin American crowd work.

\subsection{Alienation Leads To Competition} 

Our survey uncovered prevalent themes of alienation among crowdworkers. They consistently reported that their sense of isolation was not self-imposed; rather, it was primarily attributed to the platform's emphasis on rewarding efficiency rather than fostering interpersonal connections.

From this, it becomes apparent that crowdsourcing platforms, like Toloka, likely cultivate a culture that prioritizes and rewards workers' competition with one another, rather than encouraging collaboration.  Related, it is well-documented that some crowdsourcing platforms employ a deceptive tactic by simulating community promotion \cite{10.1145/3170427.3188657, rosenblat2018uberland}, only to exploit it as a means to manipulate workers into competing with each other and coercing them into performing additional labor. For example, Austin Toomb's work on Uber and Lyft's marketing campaigns found that the companies falsely advertised  feelings of a caring ethos in the hopes to gain more workers on their platform \cite{10.1145/3170427.3188657}.  This finding relates to our study as participant's reported Toloka advertising a space of support in the early onboarding period and then abandoning them once they used the platform more. Our findings also lead us to argue that the sense of isolation experienced by crowdworkers is not necessarily intrinsic to their personal feelings of not wanting to help their community, but rather a result of the algorithms optimized by crowdsourcing platforms. Toloka like many crowdsourcing platforms thrive off the workers being online and isolated from each other \cite{gray2019ghost}, thus being more alienated.  While these platforms offer benefits such as flexible scheduling and autonomy, they may also undermine connections among workers to discourage collective organizing. 
The crowd work economy has been widely discussed by many scholars within HCI  community. Research spans from critiques on crowd work, to the need to create tools to help crowdworkers. The use of digital resistance has been used to help organizers in the online community of crowd work give complaints and speak their minds.  The website We are Dynamo was a collective action campaign that gave crowdworkers on Amazon Mechanical Turk the opportunity to organize against resisting Amazon's exploitative nature of crowd work.\cite{10.1145/2702123.2702508} This site was deployed in 2011 however has not been monitored and is currently no longer online. This work compares to other crowd work tools such as Turkopticon. \cite{10.1145/2470654.2470742}  where this tool allows workers to write reviews on requester on Amazon Mechanical Turk, rating them on their communication, fairness, promptness, and generosity. This application empowers workers by allowing them to choose tasks before accepting them, addressing issues of dishonest requesters that often result in unfair treatment and pay.

To have such growning scholarship dedicated to giving more right to crowdworkers translates to the \cite{10.1145/3197391.3197394}  lack of emphasis on community building further reinforces this notion of competition and alienation. Workers, as indicated by our findings, want  better communication channels for connection, however the platform's design and priorities are likely hindering their requests. In agreement with prior work, our findings support the need for both crowd platforms to value worker's desire for community as well as independent platforms to empower workers to communicate about their labor with one another. As seen in workshops and research around the  need for design and optimizing algorithms with a caring ethos we argue that crowdsourcing platforms would benefit greatly due to workers seeing each other as equal rather than competition. The reluctance to foster meaningful connections and encourage diverse forms of individuality can be linked to neocolonialist ideologies \cite{posada2021coloniality, alvarado2021decolonial}. 

\subsection{Digital Colonialism Found in Crowd Work}

Digital colonialism is a neo-colonial idea where vulnerable populations are used in exchange for their data and labor \cite{kwet2019digital, coleman2018digital}. Our research highlights characteristics of digital colonialism, within crowdworker from Latin America predominantly comprised middle and lower-class individuals performing tasks for a European-based digital labor platform \cite{posada2021coloniality}. For example, while many participants desired connection, we also observed a growing number of participants voluntarily alienating with one another in the hopes of receiving higher rewards (in other words, giving in to the colonial values prioritized by the platform's design). According to social theorist Frantz Fanon \cite{alma9952116552801401}, individualism is a core element of colonialism and must be eradicated in the process of decolonization:
\begin{quote}
\textit{"The native intellectual had learnt from his masters that the individual ought to express himself fully. The colonialist bourgeoisie had hammered into the native's mind the idea of a society of individuals where each person shuts himself up in his own subjectivity, and whose only wealth is individual thought." \cite{alma9952116552801401}}  
\end{quote}

Considering our findings, we can draw a parallel between Fannon's notion of the "native mind" with  crowdworkers, while the bourgeoisie represents the crowd-sourcing platforms. These platforms are intentionally structured to foster competition and prioritize individual achievement \cite{gray2019ghost}, which in turn hinders the willingness of crowdworkers to collaborate collectively. When considering the ways in which crowdsourcing markets might embody colonial principles, it becomes essential to also highlight the six dimensions of power that underlie data colonialism. These dimensions encompass the ``economy'', ``political systems'', ``knowledge production'', ``personal experiences'', ``the environment'', and the intricate interplay of ``social and technological systems''. All of these aspects play a crucial role in understanding the various forms of data extraction and possible colonialism within crowd work. Considering these aspects, it becomes essential to acknowledge that in the context of the crowdsourcing economy, the success of platforms relies on the data contributed by workers, as opposed to the requesters: ``\textit{Data centered economies foster extractive models of resource exploitation, the violation of human rights, cultural exclusion, and ecocide. Data extractivism assumes that everything is a data source}'' \cite{doi:10.1177/1527476419831640}. In the context of crowdsourcing, the primary source of data comes from the labor and information provided by workers to requesters \cite{gray2019ghost}. These requesters compensate workers for their often time-consuming tasks, sometimes offering minimal payment
\cite{10.5555/AAI29394279}. Our survey revealed that workers engaged in tasks like translation (13\%), labeling data (35\%), Beta Testing (25\%) , Survey (25\%). A significant portion of their tasks involved sharing their rich cultural information, which could entail completing surveys or offering their insights for business-related objectives.However, when extracting data from these workers, requesters often fail to acknowledge the disproportionately low wages they offer for substantial studies and tasks. Our survey revealed an average daily compensation of less than \$10 for the workers' personal information.  This data extraction, connects to data colonialism, especially due to the collection and storage of information used to benefit of industry actors located in the global north, with the crowdsourcing companies capitalizing workers cultural knowledge \cite{kwet2019digital, coleman2018digital}. This lopsided power dynamic warrants thorough investigation, particularly as crowdsourcing continues to expand on a global scale.

Numerous researchers, including Nick Couldry, Ulises Mejias, Milan Stefania, and  Robert Kitchin \cite{alma9951997808401401,doi:10.1177/1527476419837739,alma9952096709201401} , have discussed  strategies to resist forms of data colonialism. They emphasize the necessity for introspection into datafication, which is the process of converting real-world information into digital data, enabling analysis and processing \cite{alma9951997808401401,doi:10.1177/1527476419837739,alma9952096709201401}. Datafication is driven by digital technologies and sensors , facilitating data-driven insights and decision-making in various fields \cite{MejiasUlisesA2019D}. Recognizing datafication's significant influence on societal decision-making \cite{alma9952212351401401}, researchers emphasize the importance of scrutinizing data sources, especially in societies pursuing equality. This scrutiny becomes crucial in contexts with fragile democracies and limited regulations \cite{alma9952212351401401}, especially as certain populations (e.g., crowdworkers) may face harm and adversity. For example, the prominent work of \cite {doi:10.1177/1527476419831640}, argues that: \textit{"Colonialism and neocolonialism should be understood in terms of their effects on power relations across borders and their reproduction
by technocrats, universities, and governments within borders of colonized countries"} \cite {doi:10.1177/1527476419831640}. Given this, it becomes pivotal to comprehend the resistance to datafication in Global South countries. HCI researchers must grapple with the connections between colonization's historical impact and its reverberations in the digital age. Part of the solution can be to consider the design of culturally aware tools that could support crowdworkers in the global south \cite{reynolds2023cima}.  When designing culturally sensitive tools for crowdworkers, it can be beneficial to consider HCI research on decolonization \cite{wong2020reflections,lazem2021challenges,alvarado2021decolonial}. This research may provide additional insights for designing crowd work tools, such as ``prioritizing workers' personal experiences over strict categorization'' or ``being aware of structural factors impacting their well-being'' \cite{pendse2022treatment}. Decolonized tools for crowdworkers can be designed to reduce power imbalances and empower workers to control their well-being \cite{karusala2021future}. The presence of digital colonialism in crowd work motivates us to focus on the Latin American and Caribbean regions. Studying data colonialism in Latin America is essential for understanding power imbalances in the global data economy \cite{reynolds2023cima}. Latin America often plays a peripheral role in this landscape, with data extraction and exploitation by more dominant global actors \cite{Medina_2023}. Examining data colonialism can shed light on these dynamics and their implications. Studying data colonialism can also help to protect data sovereignty in the region \cite{MejiasUlisesA2019D,alma9951997808401401,doi:10.1177/1527476419837739,alma9952096709201401}. Notice that data sovereignty refers to the concept that data is subject to the laws and governance of the country or region in which it is located or originates \cite{doi:10.1177/1527476419837739}. By studying and understanding current data colonialism, governments, researchers, workers, and practitioners in Latin America can develop strategies to assert greater control over their data. Studying data colonialism within Latin America can also help to address social inequalities , promote ethical data practices , shape policies , and contribute to global discussions on data exploitation \cite{alma9952212351401401}. Moreover, we draw upon this research of data colonialism to shape our own investigations, aiming to gain insights into the opportunities and challenges faced by workers in the Global South. This comprehension holds immense importance as we navigate the dynamic terrain of technology, influence, and equity.

\subsection{Gender and Crowd Work}
In this section we present our discussion on how gender was affected within crowdwork. We begin with how participants felt about their gender identity affecting their relationship with crowdwork and then explore the recommendations we concluded from these results.

\subsubsection{Gender Neutrality Pushed on Workers}

We observed that participants did not believe gender affected their day-to-day routine on the Toloka platform. This is surprising given the growing reports of pay gaps and inequalities that affect women in digital labor marketplaces \cite{10.1145/2998181.2998327, Posch_2022}. We support our participants feelings about gender neutrality within their workplace. However, it is important to take into account that growing studies around crowd work, have shown that women often make up majority of the crowdsourcing market \cite{Posch_2022}, and the algorithms on crowdsourcing platforms often undervalue women, resulting in female crowdworkers being assigned fewer tasks \cite{10.1145/2998181.2998327,19e06a23-76eb-37db-beab-6bd8b88aac98, 10.1145/3491102.3501834}, and in some cases, even receiving reduced payments as a result  \cite{10.1145/2998181.2998327}. From this, we contend that while workers may perceive Toloka as fair and egalitarian, the algorithm itself is not necessarily neutral. Therefore, the impact of personal identity and gender may not be readily visible to crowdworkers.
 
Recent research has delved into the opacity of algorithms, shedding light on how they can obscure perceptions of fairness, creating an illusion of neutrality \cite{buolamwini2018gender,10.1145/3442188.3445885}. This area of study has gained prominence, particularly in the realm of fair AI. We, as researchers, acknowledge that algorithms are not neutral. Drawing from the work of notable scholars like Joy Buolamwini, Ruha Benjamin, Safia Noble, Cathy O'Neil, and others, it becomes evident that biased datasets and biased design standards can contribute to biased job recommendation systems within crowdsourcing platforms \cite{10.1145/2998181.2998327}. This, in turn, has the potential to perpetuate harm and deepen gender, class, and racial disparities \cite{buolamwini2018gender, alma9952212351401401}. We emphasize the importance of conducting audits on crowdsourcing platforms to gain a deeper critique of the crowdsourcing markets \cite{sandvig2014auditing}. On another note, considering the way workers perceive crowdsourcing markets as neutral, there is likely value in educating them about potential disadvantages tied to their backgrounds. This understanding can pave the way for effective interventions and support systems for workers impacted by biased algorithms and interfaces. For instance, if workers remain unaware of limitations stemming from their gender, they may not see the need for tools specifically tailored for them \cite{reynolds2023cima}. In this space, it is also important to acknowledge that observed pay disparities between men and women crowdworkers could also be attributed to women having other responsibilities and not being able to dedicate as much time to the work \cite{10.1145/3476060}. Providing workers with insights into the multifaceted factors contributing to disparities in treatment and payment can empower them to seek appropriate support.

\subsubsection{Workers Want Connection}

In light of the heightened manifestations of alienation engendered by crowdsourcing platforms \cite{gray2019ghost}, we also see that crowdworkers have articulated grievances concerning Toloka's approach to their technical needs. Our observations underscore that participants exhibited a liking for engendering interconnections within the cohort of their peers as represented in \ref{fig:genderstats}.B. Our findings shows that  66\% of women crowdworkers agreed with wanting to have increased online connection and community. A smaller yet significant portion of men in our sample (50\%) also expressed a desire for such connections. Interestingly, as depicted in Figure \ref{fig:genderstats}.C, women crowdworkers used a wider variety of tools and online platforms to seek connections compared to men. From this, we argue for the creation of digital tools dedicated to helping Latin American crowdworkers foster online communities.  This tool we beleive could aim in empowering crowdworkers to connect with each other by enabling them to share recommendations, offer advice, and collaborate, thereby collectively addressing the prevalent isolation issues encountered in crowd work.  Educating and informing crowdworkers about these tools is likely also crucial, particularly as Figure \ref{fig:genderstats}.C highlights that only a small number of workers currently employ tools for community building, despite their expressed desire for connection building (Fig. \ref{fig:genderstats}B). Furthermore, as shown in Figure \ref{fig:genderstats}C, men tend to use different online spaces for community building compared to women. A formative study could help determine if gender-specific community building tools are warranted.

\section {Limitations} 

\raggedright

Our study focuses on the experiences of crowdworkers within the Latin American and Caribbean Region.  We wanted to focus our work within this region to bring more exposure to HCI research beyond a Global North Western context. From this, we believe future work regarding crowd work needs to be explored in other places other than the Global North and specifically countries in Latin America and the Caribbean.  We argue that not all research and tool building is one size fits all and believe that expanding the HCI community to places such as Latin America and the Caribbean creates more room for post-colonial system building. Regarding the population we studied, we do not hold a monolithic viewpoint of Latin America and the Caribbean and believe that although they share some commonalities based on location, cultural norms,  ethnicity, and lived experiences, the countries remain different. Based on our findings, we also recommend conducting more in-depth research with a specific emphasis on countries where crowdworker participation is notably higher, such as Venezuela, Brazil, and Colombia. Additionally, it can be essential to investigate the factors contributing to lower crowdworker numbers in certain countries. Are specific digital skills necessary for individuals in these Latin American nations to enter the crowdsourcing markets? Or could they be participating in other crowdsourcing platforms that were not within the scope of our study? Further exploration of these phenomena is warranted.  We also believe conducting an interview study forcrowdworkers within Latin America would create more verbose research around the topic of crowdsourcing and the Global South. Our findings do not aim at statistical significance but aim at exposing and bringing more attention to the views around crowd work with Latin America and the Caribbean in hopes of exposing HCI researchers to the issues at hand with crowdsourcing in the region. Furthermore, we advocate for a broader global perspective in future research on crowdsourcing. Historically, much of the research has concentrated on workers from the United States and India. We propose that future studies delve moe into the crowdsourcing dynamics of countries in Africa, which also have a significant population of crowdworkers, as well as those in Latin America. Expanding the geographical scope of research will provide a more comprehensive understanding of crowdsourcing worldwide. We also believe that further investigations could focus on conducting interviews with participants from these regions to understand in depth their relationship with crowd work.

\section{Conclusion} 
In this paper, we employed a human-centered design approach to conduct a survey, focusing on the study of crowdworkers in Latin America and The Caribbean for the very first time. From our study, we found that some crowdworkers viewed their work as a way to achieve financial and professional independence. Surprisingly, even though they desired more social connection, these workers often felt isolated from their peers and had doubts about the quality of others' work. They were hesitant to collaborate and preferred tools that did not favor one gender over the other. Our research contributes to a better understanding of crowdwork in Latin America and the Caribbean within the field of Human-Computer Interaction (HCI). It provides insights for creating digital tools that support the region's resistance efforts. Through our work, we aim to strengthen the relationship between HCI and Latin America, similar to previous HCI research initiatives \cite{reynolds2023cima}.

%workers experienced feelings of alienation while on the platform which then in turn can bred competition between the workers. We also discovered that participants believed the platform was neutral when it came to addressing gender but soon later found women and men having different experiences on  Toloka as well as women wanting more ways to form connections. It is important that more research in building sustainable tools will enable better experiences within the crowdsourcing labor market. From this research, we saw the need to develop tools for women crowdworkers in Latin America who strive for more connection online. From our work, we hope to build bridges between HCI and Latin America similar to prior CHI research 

\bibliographystyle{ACM-Reference-Format}
\bibliography{newBib}


\end{document}